\documentclass[11pt]{article}
\usepackage{moriond,epsfig}

\def\be{\begin{equation}}
\def\ee{\end{equation}}
\def\bea{\begin{eqnarray}}
\def\eea{\end{eqnarray}}
\begin{document}
\vspace*{4cm}
\title{COLLECTIVE FLAVOUR TRANSITIONS OF SUPERNOVA NEUTRINOS}

\author{IRENE TAMBORRA$^{1,2}$}
%\footnote[1]{Speaker} }

\address{$^1$Dipartimento Interateneo di Fisica ``Michelangelo Merlin", Via Amendola 173, 70126   Bari, Italy  \\
		 $^2$Istituto Nazionale di Fisica Nucleare, Sezione di Bari, Via Orabona 4, 70126   Bari, Italy}

\maketitle\abstracts{When the neutrino density is very high, as in core-collapse supernovae, neutrino-neutrino interactions are not negligible 
and can appreciably affect the evolution of flavour. The physics of these phenomena is briefly highlighted, and their effects are shown on 
observable energy spectra from a future galactic supernova within $2 \nu$ and $3 \nu$ frameworks. Detection of such effects could provide a handle on two unknowns: the neutrino 
mass hierarchy, and the mixing angle $\theta_{13}$.}

%%%%%%%%%%%%%%%%%%%%INTRODUCTION%%%%%%%%%%%%%%%%%%%%%%%%%%%%%%%%%%%%%%%%%%%%

\section{Introduction}
%%\subsection{Producing the Hard Copy}\label{subsec:prod}
Neutrino flavour eigenstates ($\nu_e$, $\nu_\mu$, $\nu_\tau$) are related to mass eigenstates ($\nu_1$, $\nu_2$, $\nu_3$) by means of an unitary matrix $U$, 
espressed in terms of three  mixing angles $\theta_{ij}$ and  one phase $\delta$ associated to possible CP violation.

We know  rather precisely  two squared mass differences ($\delta m^2$ and $\Delta m^2$, with $\delta m^2 \ll \Delta m^2$) and two mixing angles ($\theta_{12}$ and
$\theta_{23}$). However we do not know yet the sign of $\Delta m^2$ (i.e., if the mass hierarchy is normal or inverted), nor the value of $\theta_{13}$ or of $\delta$.
 Some hints about the mass hierarchy and
the mixing angle $\theta_{13}$ could come from future core-collapse supernova events in our galaxy (estimated to occur at a rate of a few per century).

In ordinary matter, neutrinos of all flavours are subject to neutral current   interactions, whereas
 $\nu_e$'s are also subject to charged current interactions  on electrons. The $\nu_e-\nu_{\mu, \tau}$ interaction 
 energy difference is described by the Mikheev-Smirnov-Wolfestein (MSW) matter potential
%........................................................
\begin{equation}
\lambda(r) = \sqrt{2}\, G_F\, N_e(r)\ , \label{lambda}
\end{equation}
%.........................................................
where $N_e(r)$ is the electron number density; see~\cite{Pant1} for a review.

When the neutrino density is very high, as in core-collapse supernovae, $\nu-\nu$ forward scattering may also become important ~\cite{Pant}.
Such interactions induce  large,  non-linear and collective flavour conversions. Since  neutrinos of different flavours  are coupled   during their evolution history,   collective effects are very different from neutrino oscillations in matter, and  they are described by means of
the self-interaction potential $\mu$. For this purpose, for each species $\nu_\alpha$,  it is useful to introduce the 
effective density $n_\alpha$ per unit of volume and of energy \cite{Semi}. After energy integration, we get the total effective density of $\nu$ 
 ($N=N_e+N_\mu+N_\tau$) and of $\bar{\nu}$
($\overline{N} = \overline{N}_e+\overline{N}_\mu+\overline{N}_\tau$) per unit volume. The potential $\mu$,  at any radius $r$, reads 
%........................................................
\begin{equation}
\mu(r) = \sqrt{2}\, G_F\, [N(r)+\overline{N}(r)]\ . \label{mu}
\end{equation}
%.........................................................

\begin{figure}[t]
\centering
\label{fig1}
\vspace*{-7mm}
\hspace*{2mm}
\epsfig{figure=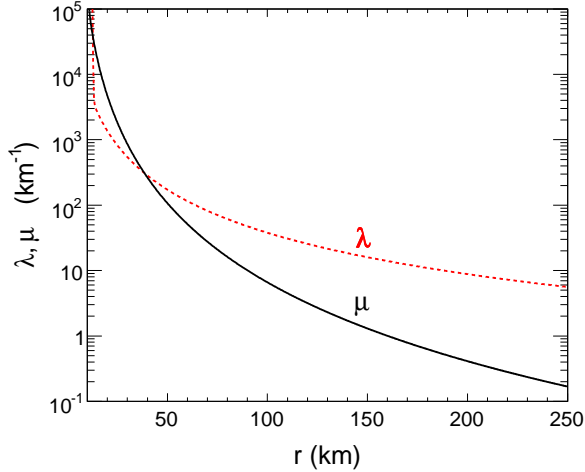,width =.57\columnwidth,angle=0}
\vspace{-4mm}
 \caption{Radial profiles adopted for the matter  ($\lambda$) and self-interaction ($\mu$) 
 potentials, in the range $r\in[10,\,250]$~km, at $t = 5$~s after the core-bounce.}
\end{figure}

Figure~\ref{fig1} \cite{Tamb} shows the matter and self-interaction potentials ($\lambda$ and $\mu$), at a representative time $t = 5$~s after the core-bounce,    for a supernova spherically-symmetric bulb model \cite{Semi} with a neutrinosphere radius $R_\nu=10$~km.

%%%%%%%%%%%%%%%%%%%%%%%%%%% FIGURE 1 %%%%%%%%%%%%%%%%%%%%%%%%%%%%%%%%
%%%%%%%%%%%%%%%%%%%%%%%%%%%%%%%%%%%%%%%%%%%%%%%%%%%%%%%%%%%%%%%%%%%%%%

The typical range for the vacuum oscillation frequency $\omega_H = \Delta m^2/2 E$ is $\omega_H \in [0.1, 5.1]$~km$^{-1}$ for $\Delta m^2 =2\times 10^{-3}$~eV$^2$. Therefore, at small radii ($r \in [10, 200]$~km) self-interactions are not negligible ($\mu \sim \omega$).  Usual MSW effects  take place
later (when $\lambda \sim \omega$) and, finally, vacuum mixing must also be considered. 

In our reference supernova scenario \cite{Schir}, we assume a galactic core-collapse supernova releasing a binding energy $E_B=3\times 10^{53}$~erg,
equally distributed among the six neutrino and antineutrino species, with luminosity decreasing 
with a time constant $\tau=3$~s. The unoscillated flux of the neutrino species $\nu_\alpha$, 
per unit of area, time and energy, is then
%...................................................................
\begin{equation}
\label{flux}
F^0_\alpha(E,\,t) = \frac{e^{-t/\tau}}{\tau}\frac{E_B}{24\pi R_\nu^2}\,\frac{\phi^0_\alpha(E)}{\langle E_\alpha\rangle}\ ,
\end{equation}
%...................................................................
where we assume normalized thermal energy spectra $\phi^0_\alpha(E)$ \cite{Tamb,Foglilast,Tamb1} 
with average energy $\langle E_\alpha \rangle$. The numerical values used for the mean energies are $\langle E_e \rangle = 10$~MeV, $\langle E_{\mu, \tau} \rangle = \langle \overline{E}_{\mu, \tau} \rangle = 24$~MeV, $\langle \overline{E}_e \rangle = 15$~MeV, where the bar labels antineutrinos.

%%%%%%%%%%%%%%%%%%%%%%%%%%%%%%%%%%%%%%%%%%%%%%%%%%%%%%%%%%%%%%%%%%%%%%
%%%%%%%%%%%%%%%%%%%%SECTION_1%%%%%%%%%%%%%%%%%%%%%%%%%%%%%%%%%%%%%%%%%%%%

\section{Collective effects in a two-flavour scenario}

In a core-collapse supernova, because of the typical neutrino energies [$E \sim O(10)$~MeV], $\nu_\mu$ and $\nu_\tau$ are both  below the threshold for $\mu$ and $\tau$ production and  have the same interactions.     
In a two-flavour approximation, we can neglect the small mass difference $\delta m^2$ and consider an effective two-family $(\nu_e,\nu_x)$ scenario 
where $\nu_x$ is either $\nu_\mu$ or $\nu_\tau$, and there are only two mass-mixing parameters, $\Delta m^2 =2\times 10^{-3}$~eV$^2$ and $\sin^2 \theta_{13} <$ few $\%$. We set $\sin^2 \theta_{13} = 10^{-4}$, but the precise value is not very important in this context. 

In the flavour basis, the neutrino system  is described by a $2 \times 2$ density matrix for each energy mode. Decomposing the density matrix over the Pauli matrices, the evolution of the system can be explained in terms of the Bloch vectors $\mathbf{P}$  and $\overline{\mathbf{P}}$ ($|\mathbf{P}| = |\overline{\mathbf{P}}| = 1$) , 
for $\nu$ and $\bar{\nu}$ respectively.    After trajectory averaging
(single-angle approximation \cite{Semi}),  the $\mathbf{P}$ and $\overline{\mathbf{P}}$ modes  obey  equations 
of motion which resemble precessions, 
%.....................
\begin{eqnarray}
\label{EOM1}
\dot\mathbf{P} &=&  \left(+\omega_H \mathbf{B}+ \lambda \mathbf{z}+\frac{\mu}{N+\overline N}\,{\int_0^\infty dE\, (n\, \mathbf{P} - \overline n\; \overline \mathbf{P})}\right) \times \mathbf{P}\ ,\\
\label{EOM2}
\dot{\overline\mathbf{P}} &=&  \left(-\omega_H \mathbf{B}+ \lambda \mathbf{z}+ \frac{\mu}{N+\overline N}\,{\int_0^\infty dE\, (n\, \mathbf{P} - \overline n\; \overline \mathbf{P})}\right) \times
\overline\mathbf{P}\ ,
\end{eqnarray}
%.....................
where  $\mathbf{B}$ is a three-dimensional ``magnetic field'' vector embedding $\theta_{13}$ \cite{Tamb,Foglilast}.  
For each energy mode, the third component of $\mathbf{P}$ ($P_z$) is related to the survival probability at the time $t$:
%..............
\begin{equation}
P_{\nu_e \rightarrow \nu_e}(t) = \frac{1}{2} \left(1 + \frac{P_z(t)}{P_z(0)}\right)\ .
\end{equation}
%..................

%%%%%%%%%%%%%%%%%%%%%%%%%%% FIGURE 2 %%%%%%%%%%%%%%%%%%%%%%%%%%%%%%%%
\begin{figure}[t]
\centering
\vspace*{-4mm}
%\hspace*{-9mm}
\epsfig{figure=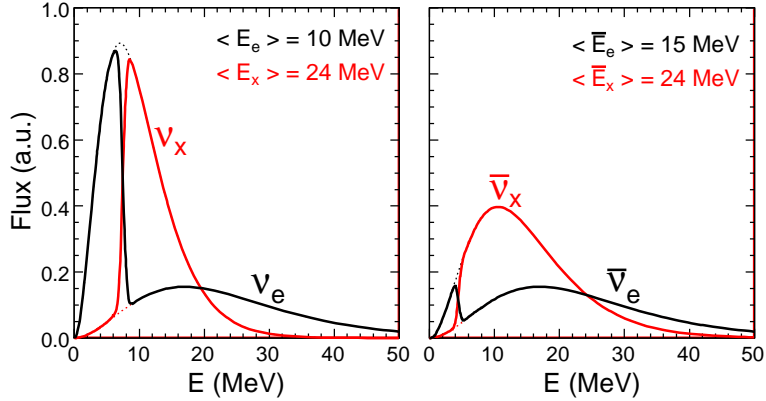,width =0.8\columnwidth}
\vspace*{-7mm}
\caption{Single-angle simulation in inverted hierarchy: 
Fluxes at the end of collective effects (at $r=250$~km, in arbitrary units) 
for different neutrino species, as a function of energy. Initial fluxes ($r=10$~km) are shown
as dotted lines to guide the eye, with average energies reported on top.
\label{fig2}}
\end{figure}
%%%%%%%%%%%%%%%%%%%%%%%%%%%%%%%%%%%%%%%%%%%%%%%%%%%%%%%%%%%%%%%%%%%%%%
Figure~\ref{fig2} \cite{Tamb} shows the fluxes $F_\alpha'$ at the end of collective effects ($r \sim$ few $\times\ 100$~km) with respect to the unoscillated ones $F_\alpha^0$.  In inverted hierarchy, a full flavour swap takes place at certain energies ($E_c \sim 7$~MeV for $\nu$ and $\overline{E}_c \sim 4$~MeV for $\bar{\nu}$, in our scenario) \cite{Tamb,Simple,Smirnov}. This full flavour conversion is called \emph{spectral split} and is an important signature of collective effects \cite{Bip1,Split1,Split2,Sigl}. It takes place only in inverted hierarchy, for any $\theta_{13} \neq 0$ \cite{Bip2}. If the hierarchy is normal or if $\theta_{13} \equiv 0$,   then $F_\alpha' \simeq F_\alpha^0$ for each $\nu_\alpha$ (i.e., there are no significant conversion effects).

%%%%%%%%%%%%%%%%%%%%%%%%%%%%%%%%%%%%%%%%%%%%%%%%%%%%%%%%%%%%%%%%%%%%%%
%%%%%%%%%%%%%%%%%%%%SECTION_2%%%%%%%%%%%%%%%%%%%%%%%%%%%%%%%%%%%%%%%%%%%%

\section{Self-interactions in a three-flavour scenario}
The two-flavour approximation captures several features of collective effects. Are these effects unchanged in a three-flavour analysis?
For this purpose, we have developed a framework with three neutrino families \cite{Tamb1}, using all three mixing angles ($\sin^2 \theta_{13} = 10^{-6}$, $\sin^2 \theta_{12} = 0.314$, $\sin^2 \theta_{23} \in [0.5, 0.36, 0.64]$), both mass differences ($\delta m^2 = 8 \times 10^{-5}$~eV$^2$ and $\Delta m^2 = 2 \times 10^{-3}$~eV$^2$) and including the one-loop $\nu_\mu-\nu_\tau$ matter potential correction \cite{Botella}. We consider the evolution at four different times ($t = 1, 5, 10, 20$~s) after the core-bounce.
%%%%%%%%%%%%%%%%%%%%%%%%%%% FIGURE 3 %%%%%%%%%%%%%%%%%%%%%%%%%%%%%%%%
\begin{figure}[t]
\centering
\vspace*{4mm}
%\hspace*{22mm}
\epsfig{figure=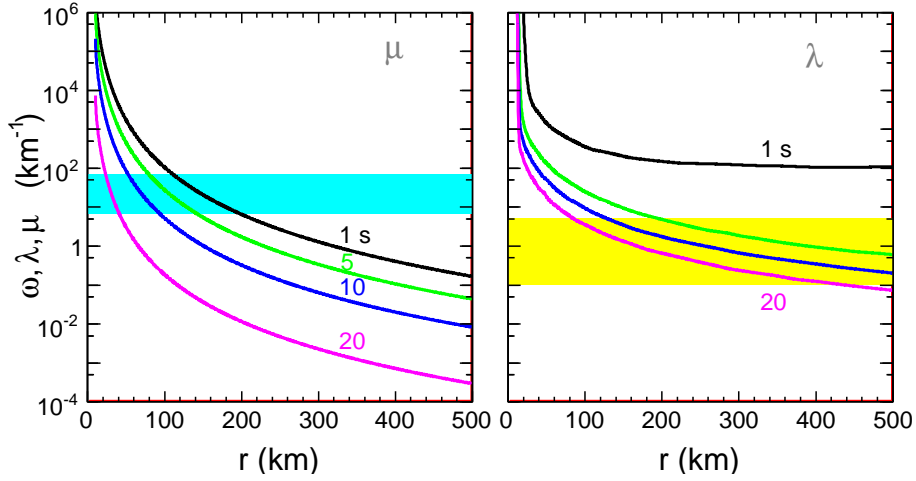,width=0.75\columnwidth}
%\vspace*{-6mm}
\caption{Radial profiles of the
self-interaction potential $\mu$ (left) and of the matter potential $\lambda$ (right). The
black, green, blue and magenta curves correspond to 
$t = 1$, 5, 10, and 20~s, respectively. In the left panel, the shaded horizontal band
marks the $\mu$ range where bipolar effects develop. In the right panel, the 
shaded band marks the range of $\omega_H$ for $E\in[1,\,50]$~MeV, where
MSW effects may develop, if the $H$-resonance condition ($\lambda \sim \omega_H$) is satisfied and if $\sin^2 \theta_{13} > 10^{-5}$. 
\label{fig3}} 
\end{figure}
%%%%%%%%%%%%%%%%%%%%%%%%%%%%%%%%%%%%%%%%%%%%%%%%%%%%%%%%%%%%%%%%%%%%%%
The self-interaction and matter potentials are plotted in Fig.~\ref{fig3}. We expect  that collective effects take place  at different radii for different $t$,
and MSW effects take place after  collective ones. In this particular case, MSW oscillations are not  relevant because of the tiny value of   $\theta_{13}$ chosen. According to this choice,  after collective effects,  we have only to consider  $\theta_{12}$ mixing to get the final $\nu$ fluxes to the Earth.

In three generations, the density operator is a $3 \times 3$ matrix in flavour basis. Decomposition over  Gell-Mann matrices provides eigth-dimensional Bloch vectors ($|\mathbf{P}| = |\overline{\mathbf{P}}| = 2/\sqrt{3}$). 
For each energy mode, the evolution equations  in inverted hierarchy are
%.........................
\begin{eqnarray}
\label{evol3nu} \dot{\mathbf{P}} &=& \left[+ (\omega_{L} \mathbf{B}_L - \omega_{H} \mathbf{B}_H)+
 \lambda\mathbf{v}  + \frac{\mu}{N+\overline N}\,{\int_0^\infty dE\, (n\, \mathbf{P} - \overline n\; \overline \mathbf{P})}\right] \times
\mathbf{P}\ ,\\
\label{evol3antinu} \dot{\overline{\mathbf{P}}} &=& \left[- (\omega_{L} \mathbf{B}_L - \omega_{H} \mathbf{B}_H)+
 \lambda\mathbf{v}  + \frac{\mu}{N+\overline N}\,{\int_0^\infty dE\, (n\, \mathbf{P} - \overline n\; \overline \mathbf{P})}\right] \times
\overline{\mathbf{P}}\ .
\end{eqnarray}
%...........................
In the above equations, the first (vacuum) terms embed the squared mass splittings via $\omega_{L,H}$ (with $\omega_L = \delta m^2/2 E$), and
the mixing angles via two ``magnetic fields'', $\mathbf{B}_{L,H}$  \cite{Tamb1}. 
The second (matter interaction) term, $\lambda \mathbf{v}$,  also includes 
the $\nu_\tau-\nu_\mu$ potential difference at one loop \cite{Botella,Esteban,Full3},  whose size is
$\delta\lambda/\lambda\simeq 5\times 10^{-5}$; the vector $ \mathbf{v}$ is a linear combination
of the unitary vectors $\mathbf{u}_3$ and  $\mathbf{u}_8$. 

In the three-flavour scenario, the survival probability is a linear combination of the third and the eighth components \cite{Das3}. In fact the analogous of the third component of $\mathbf{P}$ in the two-flavour approximation is, in the three-flavour case,  a linear combination of the third and eighth ones:
%...........................
\begin{equation}
P_z \rightarrow P_3 + \frac{P_8}{\sqrt{3}}\ .
\end{equation}
%...........................

%%%%%%%%%%%%%%%%%%%%%%%%%%% FIGURE 4-5 %%%%%%%%%%%%%%%%%%%%%%%%%%%%%%%%
\begin{figure}[t]
\begin{minipage}{18pc}
%\centering
\vspace*{-4mm}
%\hspace*{-1mm}
\epsfig{figure=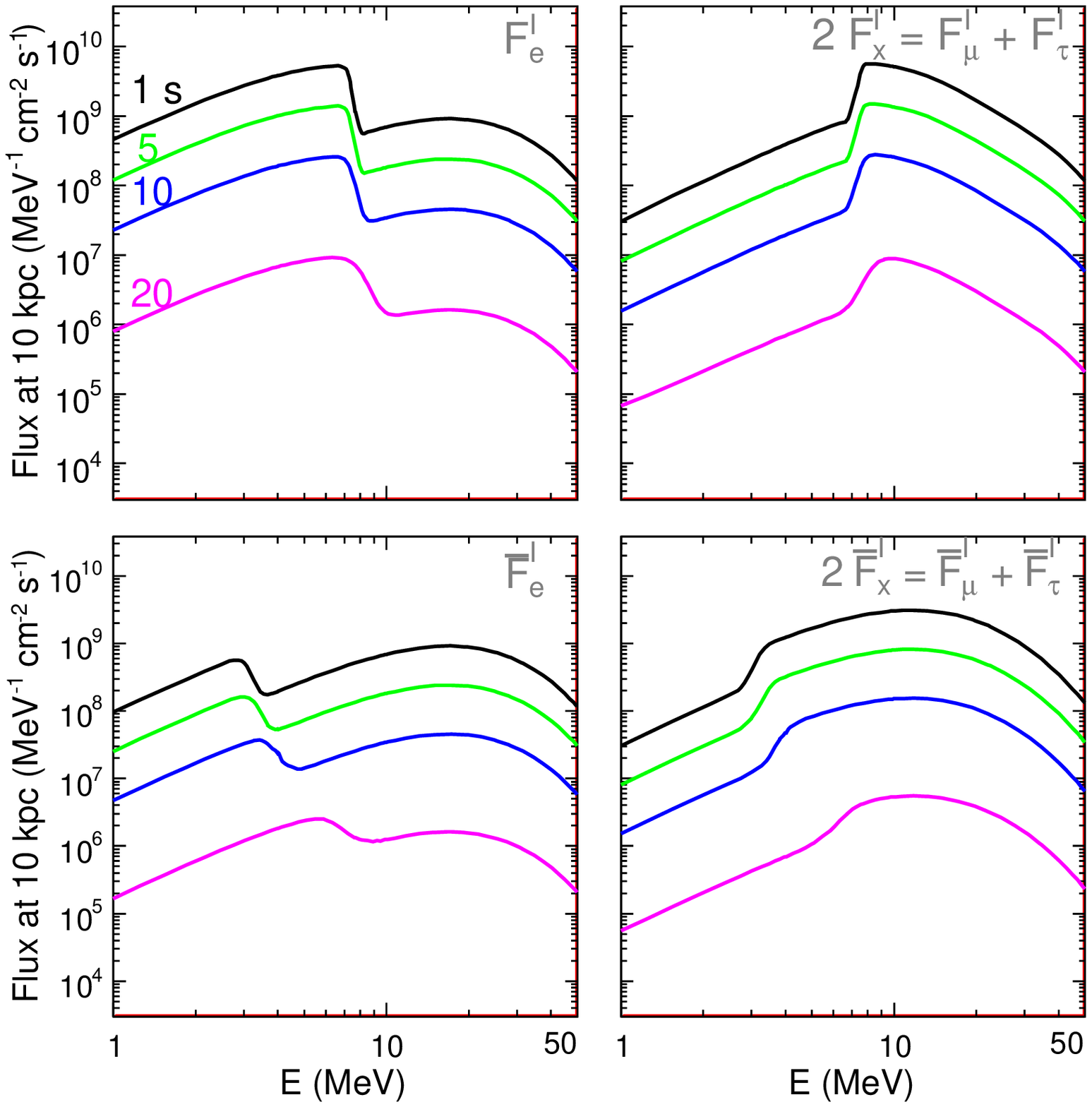,width=1.\columnwidth}
\caption{Single-angle simulation in inverted hierarchy: Fluxes of $\nu$ ($F'_\alpha$) and $\bar{\nu}$
($\overline F'_\alpha$) for four different $t$ after the core-bounce and at the end of collective effects, rescaled to $d=10$~kpc. 
\label{fig4}}\end{minipage}
\hspace{2pc}
\begin{minipage}{18pc}
%\vspace*{0mm}
%\centering
\vspace*{-3mm}
%\hspace*{92mm}
\epsfig{figure=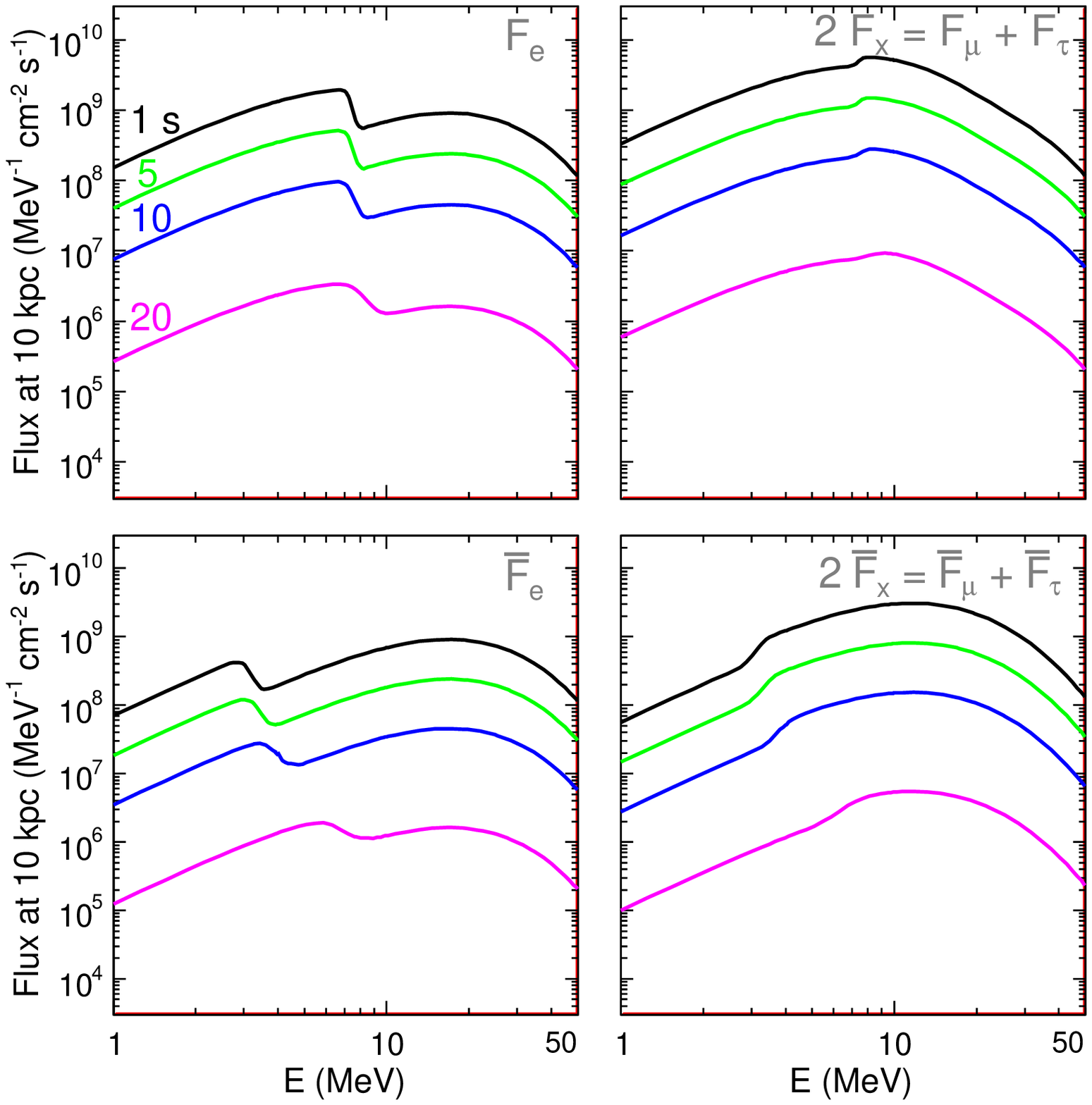,width=1.\columnwidth}
%\vspace*{0mm}
\caption{Single-angle simulation in inverted hierarchy: Final oscillated fluxes of $\nu$ ($F_\alpha$) and $\bar{\nu}$ ($\overline{F}_\alpha$) at $d = 10$~kpc (collective effects + vacuum propagation). 
\label{fig5}}
\end{minipage}
\end{figure}
%%%%%%%%%%%%%%%%%%%%%%%%%%%%%%%%%%%%%%%%%%%%%%%%%%%%%%%%%%%%%%%%%%%%%%
%%%%%%%%%%%%%%%%%%%%%%%%%%% FIGURE 6-7 %%%%%%%%%%%%%%%%%%%%%%%%%%%%%%%%
\begin{figure}[t]
\begin{minipage}{18pc}
%\centering
%\vspace*{4mm}
%\hspace*{-1mm}
\epsfig{figure=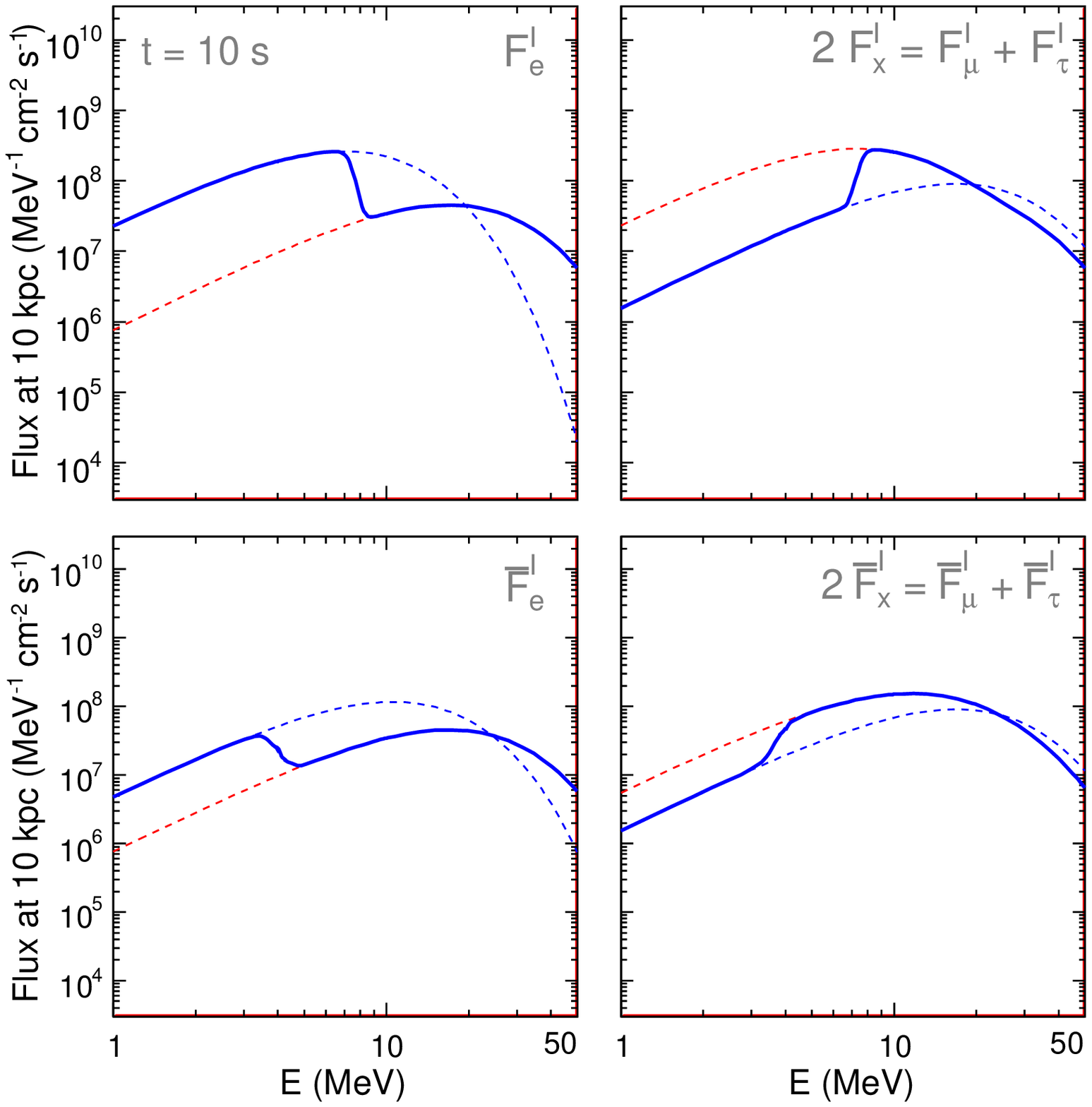,width=1.\columnwidth}
\caption{Single-angle simulation in inverted hierarchy: Fluxes of $\nu$ and $\bar{\nu}$ at the end of collective effects, for $t = 10$~s. 
Solid curves: computed fluxes $F_\alpha'$. Dashed blue and red curves: limiting behavior at low and high energies, respectively, in terms of unoscillated fluxes $F_\alpha^0$, according to Eqs.~\ref{lincomb}.
\label{fig6}}\end{minipage}
\hspace{2pc}
\begin{minipage}{18pc}
\vspace*{-7mm}
%\centering
\vspace*{6.7mm}
%\hspace*{22mm}
\epsfig{figure=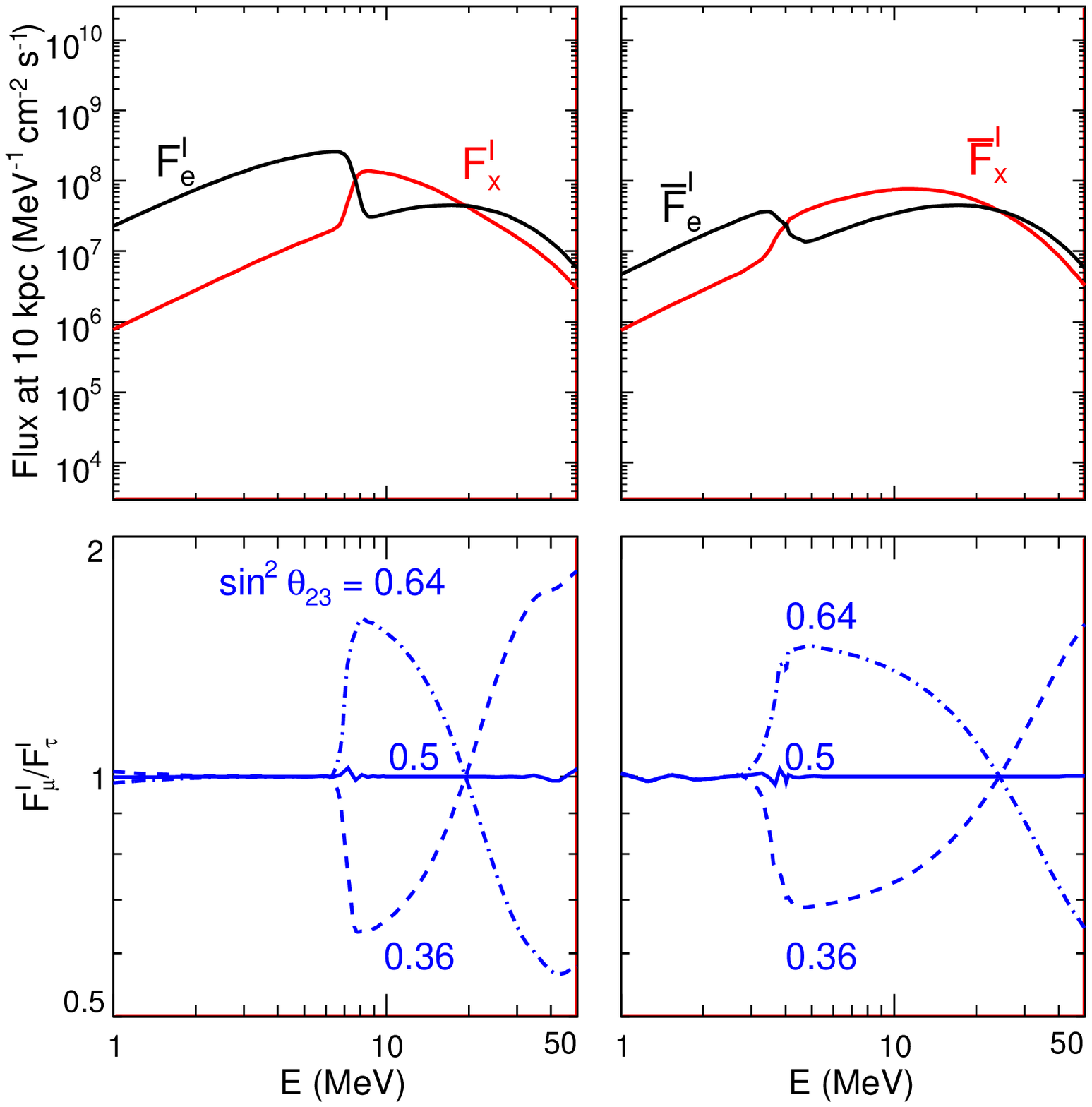,width=1.\columnwidth}
%\vspace*{0mm}
\caption{Single-angle simulation in inverted hierarchy: Neutrino and antineutrino fluxes $F'_\alpha$ at the end of collective effects ($r <500$~km), 
for $t=10$~s. Upper panels: absolute fluxes. Lower panels: muonic-to-tauonic flux ratio for $\nu$
(left) and $\overline\nu$ (right), for $\sin^2 \theta_{23} = 0.36,\, 0.5,\, 0.64$.
\label{fig7}}
\end{minipage}
\end{figure}
%%%%%%%%%%%%%%%%%%%%%%%%%%%%%%%%%%%%%%%%%%%%%%%%%%%%%%%%%%%%%%%%%%%%%%

Figure~\ref{fig4} shows the intermediate  fluxes, $F'_\alpha(E, t)$, at the end of collective effects ($r \simeq 500$~km)  for four different times after the core-bounce.  Figure~\ref{fig5} shows the corresponding fluxes $F_\alpha(E,\,t)$ at the  Earth i.e., including final vacuum mixing effects.  Split signatures of collective effects are still visible in   $\nu_e$ and $\bar{\nu}_e$ fluxes of Fig.~\ref{fig5}, and they are 
similar for different times after the core-bounce.  In view of prospective observations of galactic supernova neutrino bursts, 
the persistence of similar stepwise features for several seconds
is  useful, because   we may expect to see them also  in time-integrated spectra. The $x$-flavour split features in Fig.~\ref{fig5}  
are  suppressed  by $\theta_{12}$ mixing with respect to Fig.~\ref{fig4}.

In a two-flavour analysis a full flavour conversion ($\nu_e, \nu_x$) takes place above certain energies ($E > E_c$ for $\nu$ and $E > \overline{E}_c$ for $\bar{\nu}$),  
while in the three-flavour framework  one linear combination of $\nu_\mu$ and $\nu_\tau$ remains a ``spectator'', so that:
%.............................................................................................
\begin{equation}
\label{lincomb} 
F'_e \simeq  \left\{\begin{array}{ll}F^0_{e} & (E<E_c)\ \\ F^0_x & (E>E_c)\ \end{array} \right.~~~~~~~~~~~\mathrm{and}~~~~~~~~~~~~~~
2F'_x \simeq   \left\{\begin{array}{ll}2 F^0_{x} & (E<E_c)\ \\ F^0_e + F^0_x & (E>E_c)\ \end{array} \right.
\end{equation}
%..............................
and similarly is for antineutrinos. This limiting behavior is shown in Fig.~\ref{fig6}:  the solid curves (oscillated
fluxes) exactly superimpose to the dashed ones (linear combinations of non-oscillated fluxes, as in Eqs.~\ref{lincomb}).

Supernova neutrino fluxes are, in principle, sensitive to  the specific value of $\theta_{23}$. In fact if $\theta_{23}$ belongs to the
first (second) octant, a leading $\nu_e \rightarrow \nu_\tau$ ($\nu_e \rightarrow \nu_\mu$) takes place, or if
$\theta_{23}$ is maximal, $\nu_\mu$ and $\nu_\tau$ fluxes are exactly equal ~\cite{Esteban}. 
This  different behaviour
in $\nu_e$ conversion (in $\nu_\mu$ or in $\nu_\tau$) according to the specific value of the mixing angle $\theta_{23}$ is shown in Fig.~\ref{fig7}.  Unfortunately, $\nu_\mu$ and $\nu_\tau$ fluxes cannot be separately detected.

Although collective split signatures are similar in two and three-flavour scenarios, it is worth stressing that the former
is not a limit of the latter, because the $\nu-\nu$ interaction physics depends on the absolute luminosities 
(which are different, if shared among two or three flavours).

%%%%%%%%%%%%%%%%%%%%%%%%%%%%%%%%%%%%%%%%%%%%%%%%%%%%%%%%%%%%%%%%%%%%%%
%%%%%%%%%%%%%%%%%%%%CONCLUSIONS%%%%%%%%%%%%%%%%%%%%%%%%%%%%%%%%%%%%%%%%%%%%

\section{Conclusions}
Neutrino-neutrino interactions are not negligible when the neutrino density is very high, as in core-collapse supernovae.
$\nu-\nu$ interactions are sensitive to the mass hierarchy and $\theta_{13}$. 
In fact,  if the hierarchy is inverted and $\theta_{13} \neq 0$, split features are observable in the spectra. 
Otherwise if $\theta_{13} = 0$ or the hierarchy is normal, there are no significant conversion effects in general.

A two-flavour approximation is useful to analyze
the qualitative behavior; however  the total neutrino luminosity influences the evolution, and it changes if it is distributed on
two or three families. 
As a consequence,   three-generation analyses are important to validate the results.

%%%%%%%%%%%%%%%%%%%%%%%%%%%%%%%%%%%%%%%%%%%%%%%%%%%%%%%%%%%%%%%%%%%%%%
%%%%%%%%%%%%%%%%%%%%ACKNOWLEDGMENTS%%%%%%%%%%%%%%%%%%%%%%%%%%%%%%%%%%%%%%%%%%%%

\section*{Acknowledgments}
This work is supported in part by
the Italian  ``Istituto Nazionale di Fisica Nucleare'' (INFN) and  ``Ministero dell'Istruzione, 
dell'Universit\`a e 
della Ricerca''  (MIUR) through the ``Astroparticle Physics'' research project.  The results presented here have been
obtained in Refs. \cite{Tamb,Tamb1} in collaboration with G.L. Fogli, E. Lisi, A. Marrone, A. Mirizzi. 
I.T. is very grateful to Rencontres de Moriond EW 2009 organizers for financial support.

%%%%%%%%%%%%%%%%%%%%%%%%%%%%%%%%%%%%%%%%%%%%%%%%%%%%%%%%%%%%%%%%%%%%%%
%%%%%%%%%%%%%%%%%%%%BIBLIOGRAPHY%%%%%%%%%%%%%%%%%%%%%%%%%%%%%%%%%%%%%%%%%%%%

\end{document}